\documentclass[12pt]{article}

\PassOptionsToPackage{table}{xcolor}

\usepackage[preprint]{acl}

\usepackage{times}
\usepackage{latexsym}
\usepackage[T1]{fontenc}
\usepackage[utf8]{inputenc}
\usepackage{microtype}
\IfFileExists{inconsolata.sty}{\usepackage{inconsolata}}{}
\usepackage{graphicx}
\usepackage{booktabs}
\usepackage{dblfloatfix}   
\usepackage{enumitem}

\newcommand{\repourl}{\href{https://github.com/moorcheh-ai/memanto-benchmark}%
  {\nolinkurl{github.com/moorcheh-ai/memanto-benchmark}}}

\usepackage{pgfplots}
\pgfplotsset{compat=1.18}
\usetikzlibrary{positioning,arrows.meta,fit,backgrounds}
\usetikzlibrary{patterns}

\definecolor{cMemanto}{HTML}{C5B3EA}      
\definecolor{cMemantoEdge}{HTML}{6D4AB8}  
\definecolor{cMemzero}{HTML}{F8CBAD}      
\definecolor{cHindsight}{HTML}{C5E0B4}    
\definecolor{cHoncho}{HTML}{A9CCE3}       
\definecolor{cBuiltin}{HTML}{D9D9D9}      
\definecolor{cSupermemory}{HTML}{F4C2C2}  

\definecolor{cEdge}{gray}{0.45}  
\definecolor{cRef}{gray}{0.30}   
\definecolor{cGrid}{gray}{0.88}  

\pgfplotsset{
  dolphin base/.style={
    axis background/.style={fill=white},
    axis line style={draw=black!55},
    tick style={draw=black!55},
    tick label style={font=\footnotesize},
    label style={font=\footnotesize},
    legend style={font=\footnotesize, draw=none, fill=white, fill opacity=0.9,
                  text opacity=1, cells={anchor=west}},
    grid style={draw=cGrid},
    every node near coord/.append style={font=\footnotesize},
  },
  bar memanto/.style={fill=cMemanto, draw=cMemantoEdge, line width=0.9pt},
  bar other/.style={draw=cEdge, line width=0.4pt},
  open frame/.style={axis x line*=bottom, axis y line*=left},
  hbar labels/.style={
    nodes near coords,
    point meta=explicit symbolic,
    every node near coord/.append style={anchor=west, xshift=1pt, inner sep=1pt,
      font=\scriptsize},
  },
}
\tikzset{reference line/.style={draw=cRef, densely dashed, line width=0.6pt}}

\definecolor{memantolavender}{HTML}{DCD6F7}

\newcommand{\memantologo}{%
  \IfFileExists{figures/memanto.png}{%
    \raisebox{-0.15em}{\includegraphics[height=1.1em]{figures/memanto.png}}\hspace{0.3em}}{}}

\title{\memantologo Evaluating Memanto on DolphinBench for AI Agents}

\author{
  \textbf{Suleman Khan\textsuperscript{1,2}},
  \textbf{Rasa Rahnema\textsuperscript{2}},
  \textbf{Tara Khani\textsuperscript{2}},
  \textbf{Majid Fekri\textsuperscript{2}},
  \textbf{Fattane Zarrinkalam\textsuperscript{1}}
  \\
  \\
  \textsuperscript{1}University of Guelph,
  \textsuperscript{2}EdgeAI Innovations Inc.
  \\
  {\small \textbf{Correspondence:}
    {skhan94@uoguelph.ca},
    {rasa@edgeaiinnovations.com},
    {tara.khani@edgeaiinnovations.com},}
  \\
  {\small
    {majid.fekri@edgeaiinnovations.com},
    {fzarrink@uoguelph.ca}}
}

\begin{document}
\maketitle

\begin{abstract}
Memory benchmarks built on question answering tell an agent which fact to
recall. Workplace requests often leave that need implicit: to post an
update in the team's channel, an agent must work out which channel from
earlier interactions and use it in a tool call. We evaluate Memanto, a
memory agent that ranks memories with an information-theoretic relevance
score, on DolphinBench, which tests memory through 600 simulated
workplace tasks, each validated as requiring memory. Running in the
Hermes agent harness with GPT-6-Luna and at most 6 iterations per
conversation, Memanto completes 503 tasks (83.83\%). The strongest
configuration reported by the benchmark's authors, Mem0 with GPT-5.6-Luna
and up to 60 iterations, completes 424 (70.67\%). Memanto's run also has
the lowest total cost (\$10.86, against \$61.48 for the cheapest reported
run) and the shortest median task latency (15.32\,s, against 37.69\,s
for the fastest reported Hermes run).
Grading uses the benchmark's own judge, but the agent model differs, so
these results compare complete systems. The memory system affects both whether an agent completes
such tasks and what completion costs. Scripts and results: \repourl.
\end{abstract}

\section{Introduction}

AI agents increasingly carry out long-running work such as software
engineering, web tasks, and customer service. This work depends on
information from earlier sessions and project histories that no longer
fits in the agent's context window. Memory systems retain this
information and return it when the agent needs it
\citep{packer2023memgpt,chhikara2025mem0,rasmussen2025zep}, and memory
benchmarks measure how well they do so.

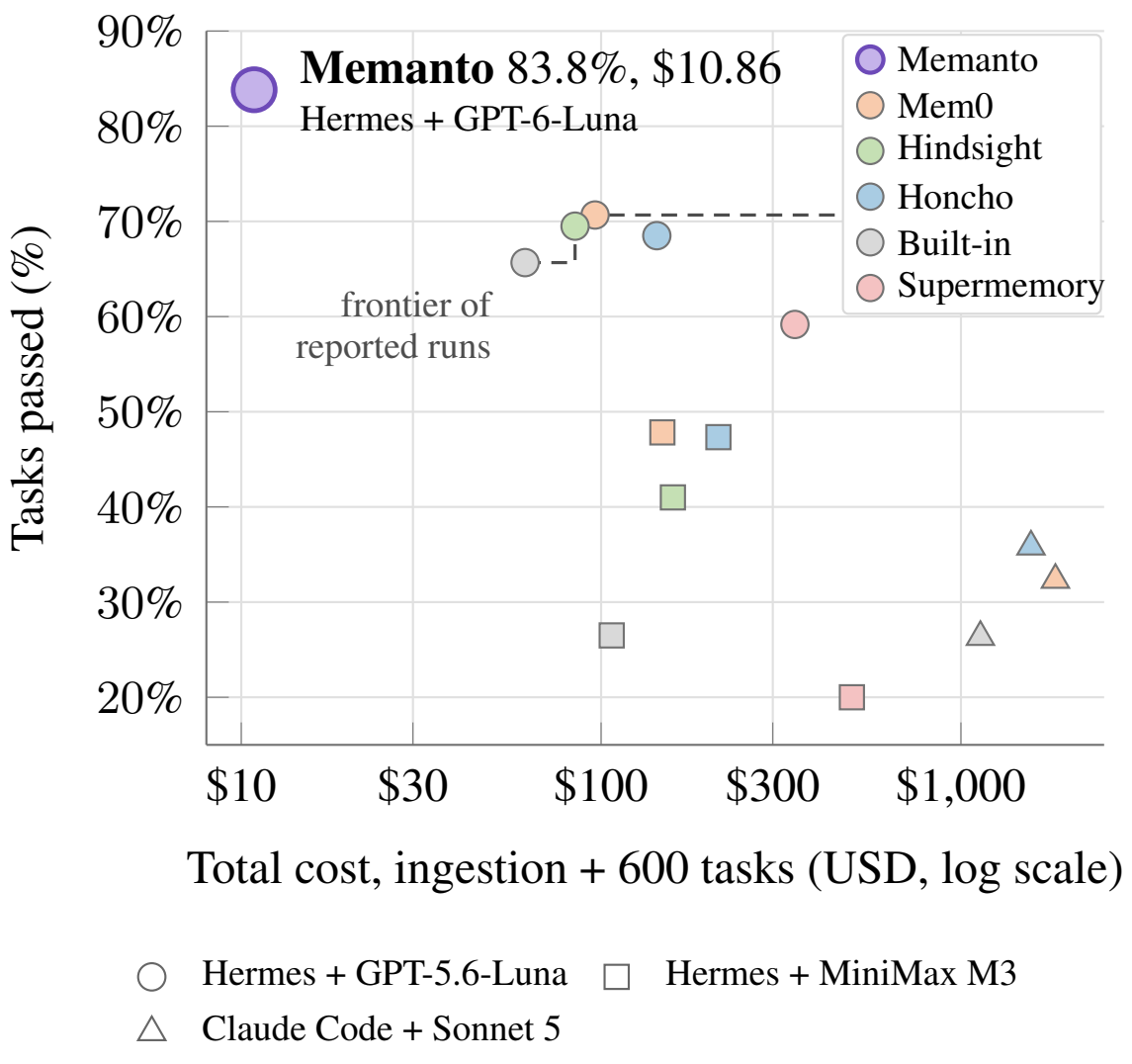
\begin{figure}[t]
  \centering
\begin{tikzpicture}
\begin{axis}[
  name=ca,
  dolphin base,
  scale only axis, width=0.78\linewidth, height=0.62\linewidth,
  xmode=log,
  xmin=8, xmax=2500,
  ymin=15, ymax=90,
  xtick={10,30,100,300,1000},
  xticklabels={\$10,\$30,\$100,\$300,\$1{,}000},
  ytick={20,30,...,90},
  yticklabel={\pgfmathprintnumber{\tick}\%},
  grid=major,
  xlabel={Total cost, ingestion + 600 tasks (USD, log scale)},
  ylabel={Tasks passed (\%)},
  open frame,
  legend columns=2,
  legend style={font=\scriptsize, at={(0.42,-0.27)}, anchor=north, column sep=5pt},
  clip=false,
]
\addplot[reference line, forget plot] coordinates {
  (61.48,65.67) (84.65,65.67) (84.65,69.50) (96.21,69.50) (96.21,70.67) (2500,70.67)};
\node[anchor=north east, font=\scriptsize, text=cRef, align=right]
  at (axis cs:57,65) {frontier of\\reported runs};

\pgfplotsset{luna/.style={only marks, mark=*, mark size=2.6pt, forget plot}}
\addplot[luna, mark options={fill=cBuiltin, draw=cEdge}]     coordinates {(61.48,65.67)};
\addplot[luna, mark options={fill=cMemzero, draw=cEdge}]     coordinates {(96.21,70.67)};
\addplot[luna, mark options={fill=cHoncho, draw=cEdge}]      coordinates {(142.85,68.50)};
\addplot[luna, mark options={fill=cHindsight, draw=cEdge}]   coordinates {(84.65,69.50)};
\addplot[luna, mark options={fill=cSupermemory, draw=cEdge}] coordinates {(345.64,59.17)};

\pgfplotsset{minimax/.style={only marks, mark=square*, mark size=2.3pt, forget plot}}
\addplot[minimax, mark options={fill=cBuiltin, draw=cEdge}]   coordinates {(107.03,26.50)};
\addplot[minimax, mark options={fill=cMemzero, draw=cEdge}]   coordinates {(148.02,47.83)};
\addplot[minimax, mark options={fill=cHoncho, draw=cEdge}]    coordinates {(211.85,47.33)};
\addplot[minimax, mark options={fill=cHindsight, draw=cEdge}] coordinates {(158.33,41.00)};
\addplot[minimax, mark options={fill=cSupermemory, draw=cEdge}] coordinates {(497.70,20.00)};

\pgfplotsset{claudecode/.style={only marks, mark=triangle*, mark size=3pt, forget plot}}
\addplot[claudecode, mark options={fill=cBuiltin, draw=cEdge}] coordinates {(1132.75,26.33)};
\addplot[claudecode, mark options={fill=cMemzero, draw=cEdge}] coordinates {(1830.57,32.33)};
\addplot[claudecode, mark options={fill=cHoncho, draw=cEdge}]  coordinates {(1565.82,35.83)};

\addplot[only marks, mark=*, mark size=4pt, forget plot,
  mark options={fill=cMemanto, draw=cMemantoEdge, line width=0.9pt}]
  coordinates {(10.86,83.83)};
\node[anchor=west, font=\footnotesize, align=left, inner sep=1pt] at (axis cs:14,83.83)
  {\textbf{Memanto} 83.8\%, \$10.86\\[-1pt]{\scriptsize Hermes + GPT-6-Luna}};

\filldraw[fill=white, draw=black!15, line width=0.4pt, rounded corners=1pt]
  (axis cs:470,60.3) rectangle (axis cs:2450,89.6);
\tikzset{every key dot/.style={circle, inner sep=0pt, minimum size=5pt, draw=cEdge},
             every key text/.style={anchor=west, font=\scriptsize, inner sep=1pt}}
\node[every key dot, fill=cMemanto, draw=cMemantoEdge, line width=0.8pt] at (axis cs:560,87) {};
\node[every key text] at (axis cs:640,87) {Memanto};
\node[every key dot, fill=cMemzero] at (axis cs:560,82.2) {};
\node[every key text] at (axis cs:640,82.2) {Mem0};
\node[every key dot, fill=cHindsight] at (axis cs:560,77.4) {};
\node[every key text] at (axis cs:640,77.4) {Hindsight};
\node[every key dot, fill=cHoncho] at (axis cs:560,72.6) {};
\node[every key text] at (axis cs:640,72.6) {Honcho};
\node[every key dot, fill=cBuiltin] at (axis cs:560,67.8) {};
\node[every key text] at (axis cs:640,67.8) {Built-in};
\node[every key dot, fill=cSupermemory] at (axis cs:560,63) {};
\node[every key text] at (axis cs:640,63) {Supermemory};

\addlegendimage{only marks, mark=*, mark size=2.6pt, mark options={fill=white, draw=black!60}}
\addlegendentry{Hermes + GPT-5.6-Luna}
\addlegendimage{only marks, mark=square*, mark size=2.3pt, mark options={fill=white, draw=black!60}}
\addlegendentry{Hermes + MiniMax M3}
\addlegendimage{only marks, mark=triangle*, mark size=3pt, mark options={fill=white, draw=black!60}}
\addlegendentry{Claude Code + Sonnet 5}
\end{axis}
\end{tikzpicture}
  \caption{Accuracy against total evaluation cost (ingestion plus 600
  tasks, log scale) on DolphinBench. Grey-outlined markers are the
  thirteen official runs: the twelve in Table~2 of
  \citet{rathi2026dolphinbench} and the Hermes, MiniMax M3, and
  Supermemory run from the official results
  \citep{dolphinbench2026results}. Marker shape
  gives the harness and agent model, and fill colour the memory system.
  The dashed staircase is the frontier of those runs. Memanto, run in
  Hermes with GPT-6-Luna, lies above and to the left of every reported
  run.}
  \label{fig:costacc}
\end{figure}

Most memory benchmarks pose questions about a long conversation history
\citep{maharana2024locomo}. Later benchmarks added temporal reasoning,
knowledge updates, and abstention \citep{wu2025longmemeval}, and
histories beyond a million tokens \citep{tavakoli2025beam}, but kept the
question-answering format. A question names what to
retrieve: ``Which messaging platform does the team use?'' tells the
system what to recall. An instruction such as ``Post an update in the
team's channel'' does not. The agent must recognize that the request
depends on an earlier interaction, retrieve the right detail, and use it
correctly in a tool call. Evaluating memory for agents therefore calls
for tasks in which the needed information must be identified, retrieved,
and acted on.

Benchmark quality matters as much as format. An audit of LoCoMo found
incorrect answers in 6.4\% of its non-adversarial questions (99 of 1,540)
\citep{penfield2026locomoaudit}, and community reports identified
arithmetic errors and mislabeled evidence in LongMemEval
\citep{longmemeval2025issues}. BEAM uses human review to check that its
questions are answerable \citep{tavakoli2025beam}, but an answerable task
need not require memory: a capable model may solve it from the request
alone.

DolphinBench \citep{rathi2026dolphinbench} addresses both concerns. Its
600 tasks are workplace actions carried out through simulated
applications, and each task is kept only if an agent passes it in two
runs when given the relevant history and fails it in two runs without
that history. Every task is therefore solvable and depends on memory.
The benchmark also requires every result to report accuracy together with
total cost and median task latency, so accuracy cannot be bought with
unreported spending or waiting time.

\begin{figure*}[t]
  \centering
\begin{tikzpicture}[
  font=\footnotesize,
  >={Stealth[length=4pt]},
  box/.style={draw=cEdge, line width=0.5pt, rounded corners=2pt, fill=white,
              align=center, inner sep=3pt, minimum height=1.05cm, text width=#1},
  box/.default=2.9cm,
  memanto/.style={box=#1, fill=cMemanto!55, draw=cMemantoEdge, line width=0.8pt},
  memanto/.default=2.9cm,
  flow/.style={->, draw=black!65, line width=0.6pt},
  lab/.style={font=\scriptsize, text=black!70, inner sep=1.5pt},
  panel/.style={draw=black!25, fill=black!3, rounded corners=4pt, inner sep=6pt},
  title/.style={font=\footnotesize\bfseries, anchor=south west, inner sep=2pt},
]
\node[box=3.1cm] (hist) at (0,0)
  {\textbf{Persona history}\\3,400--5,128 dated user messages ($\approx$500k tokens)};
\node[memanto=3.1cm, below=0.55cm of hist] (ingest)
  {\textbf{Hermes agent saves}\\one new conversation per message; Memanto \emph{remember} tool};
\node[memanto=3.1cm, below=0.55cm of ingest] (frozen)
  {\textbf{Frozen memory}\\writes confirmed, then snapshotted; read only in tests};
\draw[flow] (hist) -- (ingest);
\draw[flow] (ingest) -- (frozen);

\node[box=2.5cm, right=1.0cm of hist] (task)
  {\textbf{Task request}\\new conversation for every task};
\node[box=2.6cm, right=0.85cm of task] (agent)
  {\textbf{Hermes agent}\\GPT-6-Luna\\{\scriptsize $\leq$6 iterations per task}};
\node[box=3.3cm, right=1.0cm of agent] (tools)
  {\textbf{Simulated applications}\\Notion, Gmail, GitHub, calendar, CRM, deployment};
\node[memanto=2.6cm] (recall) at (agent |- frozen)
  {\textbf{Memanto \emph{recall}}\\ITS score $\geq 0.1$\\top $N{=}50$ memories};
\node[box=3.3cm] (grade) at (tools |- frozen)
  {\textbf{Grading}\\deterministic checks +\\LLM judge (GPT-5.6-Sol)};

\draw[flow] (task) -- (agent);
\draw[flow] ([xshift=-5pt]agent.south) -- node[lab, left, align=right] {recall\\call} ([xshift=-5pt]recall.north);
\draw[flow] ([xshift=5pt]recall.north) -- node[lab, right, align=left] {ranked\\memories} ([xshift=5pt]agent.south);
\draw[flow] ([yshift=4pt]agent.east) -- node[lab, above] {tool calls} ([yshift=4pt]tools.west);
\draw[flow] ([yshift=-4pt]tools.west) -- node[lab, below] {results} ([yshift=-4pt]agent.east);
\draw[flow] (tools) -- node[lab, right, align=left] {final state\\and trace} (grade);
\draw[flow, densely dashed] (frozen) -- node[lab, above] {read only} (recall);
\node[lab, below=1pt of grade, align=center] (rule) {pass only if every check passes};

\begin{scope}[on background layer]
  \node[panel, fit=(hist)(ingest)(frozen)(hist |- rule.south)] (p1) {};
  \node[panel, fit=(task)(agent)(tools)(recall)(grade)(rule)] (p2) {};
\end{scope}
\node[title, align=left] at (p1.north west)
  {1\enspace Ingestion\\[-1pt]{\scriptsize\normalfont once per persona}};
\node[title, align=left] at (p2.north west)
  {2\enspace Task execution\\[-1pt]{\scriptsize\normalfont once per task: 600 tasks, 200 per persona}};
\end{tikzpicture}
  \caption{How Memanto was evaluated on DolphinBench. During ingestion,
  the Hermes agent reads each dated history message in a new conversation
  and saves what it should remember through Memanto's \emph{remember}
  tool; every write is confirmed searchable, and the memory is then frozen.
  For each of the 600 tasks (200 per persona), the agent starts a new
  conversation, calls Memanto's \emph{recall} tool, which returns memories
  ranked by information-theoretic scoring (ITS), and completes the task
  through simulated applications. A task passes only when every
  deterministic and LLM-judged check passes.}
  \label{fig:workflow}
\end{figure*}
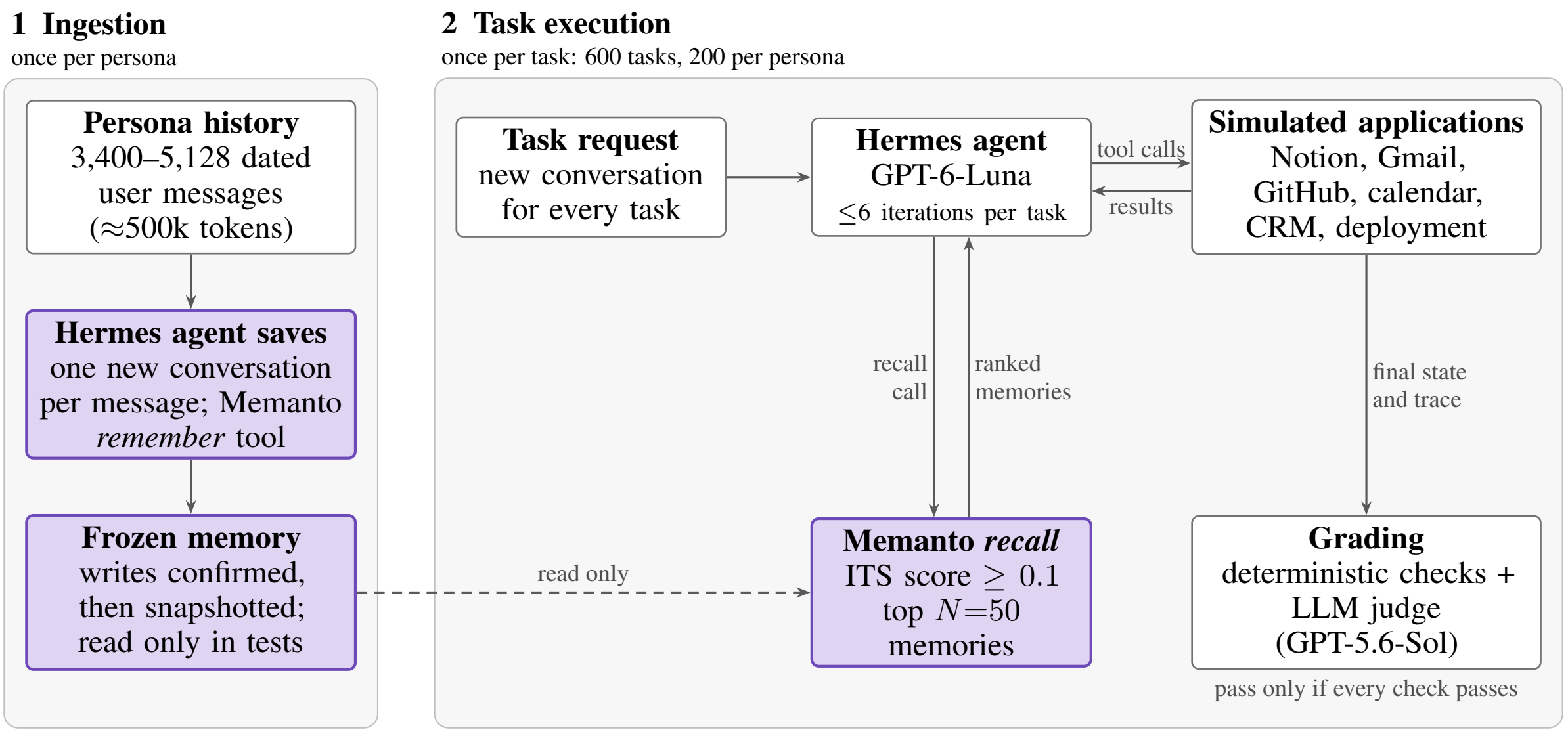

We evaluate Memanto \citep{abtahi2026memanto} on DolphinBench
(Figure~\ref{fig:workflow}). Memanto is a memory agent that selects
memories with information-theoretic scoring \citep{abtahi2026moorcheh}.
In the Hermes agent harness \citep{nous2026hermes} with GPT-6-Luna,
Memanto completes 503 of 600 tasks (83.83\%), at a total cost of \$10.86
and a median task latency of 15.32\,s (Figure~\ref{fig:costacc}). The
strongest result reported by the benchmark's authors, Mem0 in the same
harness with GPT-5.6-Luna, completes 424 tasks (70.67\%) at \$96.21 and
37.69\,s. Our contributions are:
\begin{itemize}[leftmargin=*,itemsep=1pt,topsep=3pt]
  \item a complete DolphinBench run of Memanto, reporting accuracy, cost
        split between the agent model and memory, and median latency, as
        the benchmark requires, with the scripts and results of the run
        released for reproduction\footnote{\label{fn:repo}\repourl};
  \item a per-persona comparison with the official runs, showing that
        Memanto's accuracy is stable across the three personas
        (83.0--85.0\%), including the persona every reported run found
        hardest;
  \item an analysis of how threshold-based information-theoretic
        retrieval may help when facts are revised across sessions, and of
        what a comparison with runs that differ in agent model and agent
        settings can and cannot show.
\end{itemize}

\section{Related Work}

\paragraph{Memory benchmarks.}
LoCoMo \citep{maharana2024locomo} and LongMemEval \citep{wu2025longmemeval}
evaluate memory through questions about long multi-session
conversations; LongMemEval covers information extraction, multi-session
reasoning, temporal reasoning, knowledge updates, and abstention. BEAM
\citep{tavakoli2025beam} extends conversations beyond a million tokens.
MemBench \citep{tan2025membench} and MemoryAgentBench
\citep{hu2026memoryagentbench} evaluate memory in LLM agents, the latter
by delivering information incrementally over many turns. A newer line of
work evaluates memory through agent tasks: MEMTRACK
\citep{deshpande2025memtrack} evaluates memory and state tracking across
multi-platform workplace environments, MemoryArena
\citep{he2026memoryarena} links sessions so that later subtasks depend on
earlier ones, and Mem2ActBench \citep{shen2026mem2actbench} tests whether
agents use long-term memory in task-oriented actions. DolphinBench
\citep{rathi2026dolphinbench} belongs to this line and adds per-task
validation against memory and mandatory reporting of cost and latency.

\paragraph{Memory systems for agents.}
MemGPT \citep{packer2023memgpt} manages tiers of context in the manner of
an operating system. Mem0 \citep{chhikara2025mem0} extracts and
consolidates salient facts from conversations, Zep
\citep{rasmussen2025zep} organizes memory as a temporal knowledge graph,
and Hindsight \citep{latimer2025hindsight} structures memory around
retaining, recalling, and reflecting. Honcho \citep{plasticlabs2025honcho}
and Supermemory \citep{supermemory2025} are open-source memory systems for
agents, and agent harnesses such as Hermes \citep{nous2026hermes} include
built-in memory. Memanto \citep{abtahi2026memanto} represents memories as
typed semantic records and retrieves them with information-theoretic
scoring \citep{abtahi2026moorcheh}; Section~\ref{sec:memanto} describes
the configuration we evaluate.

\section{DolphinBench}
\label{sec:dolphinbench}

DolphinBench \citep{rathi2026dolphinbench} evaluates memory through task
completion rather than question answering. It covers three knowledge-work
personas: Morgan, a startup CEO; Alex, an infrastructure engineer; and
Riley, a product manager. Each persona has a simulated history spanning
several years, with 3,400--5,128 user messages (about 500k tokens). The
histories embed changing details in everyday conversation, so an agent
must notice what is relevant, track updates, and tell current information
from outdated information.

After processing a persona's history, the agent completes 200 tasks for
that persona, 600 in total. Tasks are workplace actions carried out
through simulated applications, such as Notion, Gmail, and GitHub, and
calendar, email, CRM, and deployment tools. The task request and the
application records give the immediate context, but they do not say
which earlier interaction holds the needed detail: the agent must
recognize that memory is needed and decide what to retrieve. To validate
a task, the benchmark authors ran GPT-5.6-Luna twice with the relevant
(oracle) messages and twice without them; a task was kept only if both
runs with the messages passed every check and both runs without them
failed.

\section{Memanto}
\label{sec:memanto}

Memanto \citep{abtahi2026memanto} is a memory agent for long-horizon AI
agents. It gives an agent two tools: \emph{remember}, which saves a
memory, and \emph{recall}, which returns the memories most relevant to a
query. The agent decides when to call them. Relevance is computed with information-theoretic scoring (ITS)
from Moorcheh \citep{abtahi2026moorcheh}, which scores binarized
representations exhaustively rather than through an approximate
nearest-neighbour index, so the same query over the same memories always
returns the same ranking.

Two settings control what reaches the agent. A score threshold decides
which memories qualify, and a retrieval limit caps how many enter the
agent's context. We use a threshold of 0.1 and a limit of $N=50$. On the
ITS scale, scores below 0.1 form the lowest relevance band, labeled
irrelevant \citep{moorcheh2026docs}, so this threshold removes only
memories that ITS rates as irrelevant, while the limit bounds the size,
and therefore the cost, of the agent's context.

\begin{table*}[t]
\centering
\small
\setlength{\tabcolsep}{6.5pt}
\begin{tabular}{l rr rrr r r}
\toprule
& \multicolumn{2}{c}{\textbf{Tasks passed}}
& \multicolumn{3}{c}{\textbf{Cost (USD)}}
& \textbf{Cost per}
& \textbf{Median} \\
\cmidrule(lr){2-3}\cmidrule(lr){4-6}
\textbf{Memory system}
& \textbf{of 600} & \textbf{\%}
& \textbf{Agent} & \textbf{Memory} & \textbf{Total}
& \textbf{passed task}
& \textbf{latency (s)} \\
\midrule
\multicolumn{8}{l}{\textit{Reported by \citet{rathi2026dolphinbench}: Hermes with GPT-5.6-Luna}} \\
Built-in memory & 394 & 65.67 & 61.48 & 0.00   & 61.48  & 0.156 & 44.35 \\
Mem0            & 424 & 70.67 & 61.54 & 34.68  & 96.21  & 0.227 & 37.69 \\
Honcho          & 411 & 68.50 & 96.56 & 46.30  & 142.85 & 0.348 & 44.66 \\
Hindsight       & 417 & 69.50 & 57.99 & 26.66  & 84.65  & 0.203 & 55.31 \\
Supermemory     & 355 & 59.17 & 63.86 & 281.79 & 345.64 & 0.974 & 52.68 \\
\midrule
\multicolumn{8}{l}{\textit{This work: Hermes with GPT-6-Luna}} \\
\rowcolor{memantolavender}
Memanto & \textbf{503} & \textbf{83.83} & \textbf{2.17} & 8.69 & \textbf{10.86} & \textbf{0.022} & \textbf{15.32} \\
\bottomrule
\end{tabular}
\caption{DolphinBench results. The upper rows are the Hermes runs
reported in Table~2 of \citet{rathi2026dolphinbench}; tasks passed and cost
per passed task are derived from the reported accuracy and total cost.
The highlighted row is our run in the same harness, graded by the same
judge, with a different agent model and agent settings
(Section~\ref{sec:setup}), so
differences reflect the complete system, not the memory system alone. Cost covers ingestion and all 600
tasks; latency is the median time per task, including tool use. Bold
marks the best value in each column where Memanto's run is best.}
\label{tab:results}
\end{table*}

\section{Experimental Setup}
\label{sec:setup}

\paragraph{Systems compared.}
We ran Memanto in the Hermes agent harness \citep{nous2026hermes} with
GPT-6-Luna as the agent model. We compare it with the five memory
configurations that \citet{rathi2026dolphinbench} evaluated in the same
harness: Hermes built-in memory, Mem0 \citep{chhikara2025mem0}, Honcho
\citep{plasticlabs2025honcho}, Hindsight \citep{latimer2025hindsight},
and Supermemory \citep{supermemory2025}. Their figures are taken from
Table~2 of \citet{rathi2026dolphinbench}, where all five use GPT-5.6-Luna;
we did not re-run them. Per-persona scores and the split of agent cost
between ingestion and tests come from the official run records released
with the benchmark \citep{dolphinbench2026results}.
Figure~\ref{fig:costacc} also shows the official runs with MiniMax M3 in
Hermes and with Claude Sonnet 5 in Claude Code.

\paragraph{Agent settings.}
Our run uses medium reasoning effort and in each stage, only the tools that
stage needs: Memanto's \emph{remember} during ingestion, and the task's
simulated application with Memanto's \emph{recall} during tests.

\paragraph{Protocol.}
Our run has the two stages DolphinBench defines: the full history of a
persona is processed before any test is taken.
\emph{Ingestion:} as the official protocol requires, the Hermes agent
processes each persona's history messages in order, one new conversation
per message. Its only tool is Memanto's \emph{remember}, and an
ingestion instruction asks it to save the information in the message,
preserving dates, amounts, names, and corrections. Automatic capture is
disabled, so the memory holds only what the agent chooses to save, and
each write is confirmed to be searchable before the run continues. The
completed memory is then frozen: its contents are snapshotted and hashed,
and before testing the live memory is checked against the snapshot.
\emph{Task execution:} for each task, the agent starts a new
conversation with the task's application tools and Memanto's
\emph{recall} tool (ITS threshold 0.1, $N=50$), and completes the task
through the simulated applications. Memory writes are blocked during
tests.
Freezing memory and starting every task in a new conversation keeps
tasks independent: nothing learned while solving one task can help with
another. The official protocol requires both: a fresh conversation for
each test, and memory that tests cannot change
\citep{dolphinbench2026results}. The scripts for both stages and the
results of our run are in the accompanying
repository.\footref{fn:repo}

\paragraph{Metrics.}
We report the three measures DolphinBench requires.
\emph{Accuracy} is the percentage of the 600 tasks passed. Grading
combines deterministic checks, for requirements such as tool calls,
identifiers, dates, and values, with an LLM judge for requirements that
need semantic interpretation \citep{zheng2023judging}; a task passes only
when every check passes. As in the reported runs, the judge is
GPT-5.6-Sol, the benchmark's own, with its grading prompts and verdict
logic unchanged.
\emph{Cost} is the total spent on ingestion and on the 600 tasks, split
into agent-model tokens, as estimated by Hermes from the recorded API
usage, and memory operations, measured as the Moorcheh Cloud usage
credits the run consumed.
\emph{Latency} is the median time per task, including tool use.

\section{Results}
\label{sec:results}

\paragraph{Task completion.}
Table~\ref{tab:results} reports the results. Memanto's run passes 503 of
600 tasks (83.83\%). The strongest reported configuration, Mem0, passes
424 (70.67\%), so Memanto's run passes 79 more tasks, a difference of
13.16 percentage points. Relative to Hermes built-in memory (394 tasks,
65.67\%), it passes 109 more. Because a task passes only when every check
passes, these are differences in end-to-end task completion, not in
recall alone. The run reaches this result with at most 6 iterations per
conversation, a tenth of the 60 allowed in the reported runs: a tighter
limit leaves the agent fewer chances to recover from a wrong step, so it
works against task completion rather than for it.

\begin{figure}[t]
  \centering
\begin{tikzpicture}
\begin{axis}[
  dolphin base, open frame, hbar labels,
  scale only axis, width=0.66\linewidth, height=0.5\linewidth,
  xbar, bar width=7.5pt, bar shift=0pt,
  xmin=0, xmax=66,           
  xtick={0,10,...,60},
  xmajorgrids,
  xlabel={Median time per task (s), lower is better},
  symbolic y coords={hindsight,supermemory,honcho,builtin,memzero,memanto},
  ytick={hindsight,supermemory,honcho,builtin,memzero,memanto},
  ymin=hindsight, ymax=memanto,
  yticklabels={Hindsight, Supermemory, Honcho, Built-in, Mem0, \textbf{Memanto}},
  enlarge y limits=0.09,
  clip=false,
]
\addplot[bar memanto]                  coordinates {(15.32,memanto)     [\textbf{15.3\,s}]};
\addplot[bar other, fill=cMemzero]     coordinates {(37.69,memzero)     [37.7\,s]};
\addplot[bar other, fill=cBuiltin]     coordinates {(44.35,builtin)     [44.4\,s]};
\addplot[bar other, fill=cHoncho]      coordinates {(44.66,honcho)      [44.7\,s]};
\addplot[bar other, fill=cSupermemory] coordinates {(52.68,supermemory) [52.7\,s]};
\addplot[bar other, fill=cHindsight]   coordinates {(55.31,hindsight)   [55.3\,s]};

\draw[reference line]
  ({axis cs:32.19,hindsight}|-{rel axis cs:0,0}) --
  ({axis cs:32.19,hindsight}|-{rel axis cs:0,1})
  node[above, font=\scriptsize, text=cRef, align=center]
  {fastest reported run, any setup (32.2\,s)};
\end{axis}
\end{tikzpicture}
  \caption{Median time per task, tool use included. Comparison runs use
  Hermes with GPT-5.6-Luna \citep{rathi2026dolphinbench}; Memanto's run
  uses Hermes with GPT-6-Luna, medium reasoning effort, and at most 6
  iterations per conversation; latency is the measure most affected by
  these settings (Section~\ref{sec:discussion}). The dashed line marks
  the fastest reported run in any setup: Claude Code with Claude Sonnet~5
  and built-in memory, 32.2\,s at 26.3\% accuracy. Both reference values
  are from Table~2 of \citet{rathi2026dolphinbench}.}
  \label{fig:latency}
\end{figure}
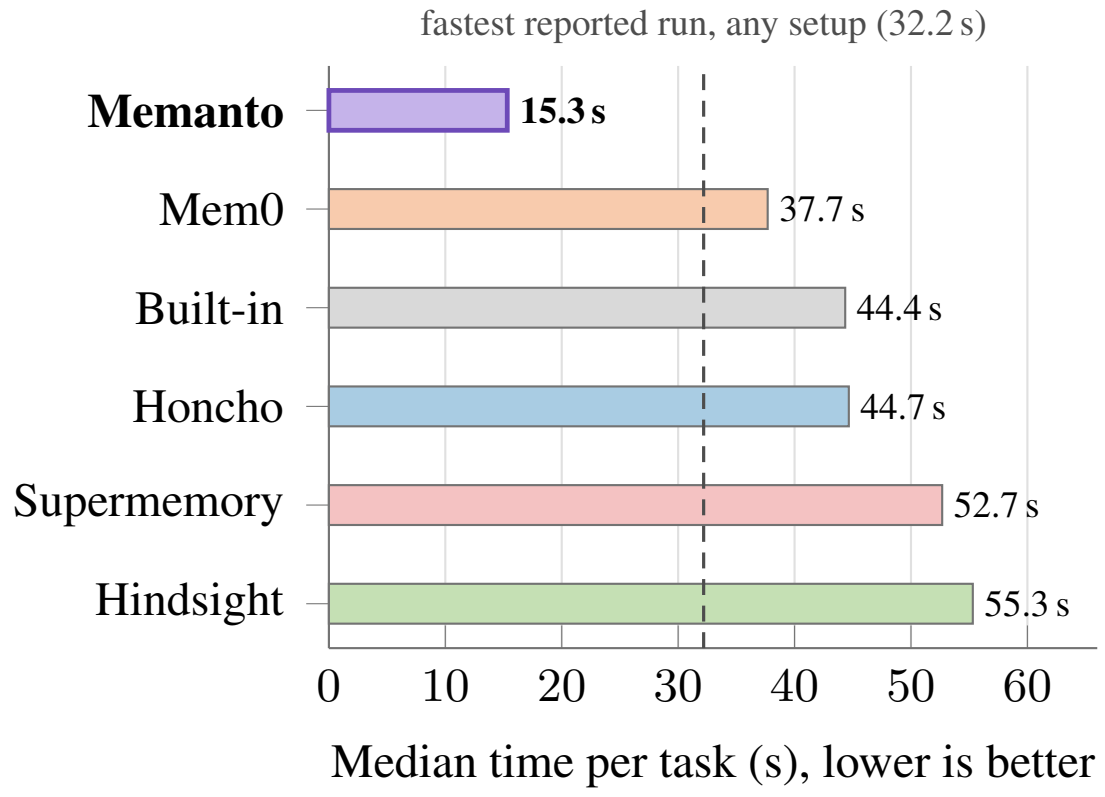

\paragraph{Cost and latency.}
Memanto's run costs \$10.86 in total: \$8.69 for memory operations and
\$2.17 for agent-model tokens. The cheapest reported run, Hermes built-in
memory, has no separate memory charge but spends \$61.48 on agent tokens;
Memanto's total is 82.3\% lower. Per passed task, Memanto's run costs
2.2 cents, against 15.6 cents for built-in memory and 22.7 cents for
Mem0. Its median task latency is 15.32\,s, against 37.69\,s for Mem0,
the fastest reported Hermes run, and 44.35\,s for built-in memory
(Figure~\ref{fig:latency}); it is also less than half the 32.19\,s of
the fastest reported run in any setup, Claude Code with built-in memory,
which passes 26.33\% of tasks. Among the reported runs, the cheapest is
neither the most accurate nor the fastest; Memanto's run leads on all
three measures at once. Part of its cost and latency advantage likely
comes from the tighter iteration limit and lower reasoning effort
(Section~\ref{sec:discussion}). Because DolphinBench measures the cost
of completing its tasks, and this run completed more of them than any
reported run, the saving is not bought with lower task completion. Figure~\ref{fig:costacc} places it among all
thirteen official runs across three harness and model setups: it is both
cheaper and more accurate than every one of them.

\begin{figure}[t]
  \centering
\begin{tikzpicture}
\begin{axis}[
  dolphin base, open frame,
  scale only axis, width=0.8\linewidth, height=0.48\linewidth,
  ybar, bar width=4.3pt,
  ymin=0, ymax=200,
  ytick={0,50,...,200},
  ymajorgrids,
  ylabel={Tasks passed (of 200)},
  symbolic x coords={Morgan,Alex,Riley},
  xtick=data,
  xticklabels={{Morgan\\[-1pt]\scriptsize startup CEO},
               {Alex\\[-1pt]\scriptsize infrastructure engineer},
               {Riley\\[-1pt]\scriptsize product manager}},
  xticklabel style={align=center, anchor=north},
  enlarge x limits=0.2,
  legend columns=3,
  legend style={at={(0.5,1.04)}, anchor=south, column sep=3pt, font=\scriptsize},
  legend image code/.code={\draw[#1] (0cm,-0.08cm) rectangle (0.2cm,0.12cm);},
]
\addplot[bar memanto,
  nodes near coords,
  every node near coord/.append style={font=\scriptsize\bfseries, text=cMemantoEdge}]
  coordinates {(Morgan,170) (Alex,166) (Riley,167)};
\addlegendentry{\textbf{Memanto}}
\addplot[bar other, fill=cMemzero]     coordinates {(Morgan,156) (Alex,132) (Riley,136)};
\addlegendentry{Mem0}
\addplot[bar other, fill=cHindsight]   coordinates {(Morgan,144) (Alex,127) (Riley,146)};
\addlegendentry{Hindsight}
\addplot[bar other, fill=cHoncho]      coordinates {(Morgan,144) (Alex,120) (Riley,147)};
\addlegendentry{Honcho}
\addplot[bar other, fill=cBuiltin]     coordinates {(Morgan,131) (Alex,111) (Riley,152)};
\addlegendentry{Built-in}
\addplot[bar other, fill=cSupermemory] coordinates {(Morgan,138) (Alex,98)  (Riley,119)};
\addlegendentry{Supermemory}
\end{axis}
\end{tikzpicture}
  \caption{Tasks passed per persona (out of 200). Comparison runs use
  Hermes with GPT-5.6-Luna; their per-persona scores are from the official
  run records \citep{dolphinbench2026results}. Memanto's run uses Hermes
  with GPT-6-Luna.}
  \label{fig:persona}
\end{figure}
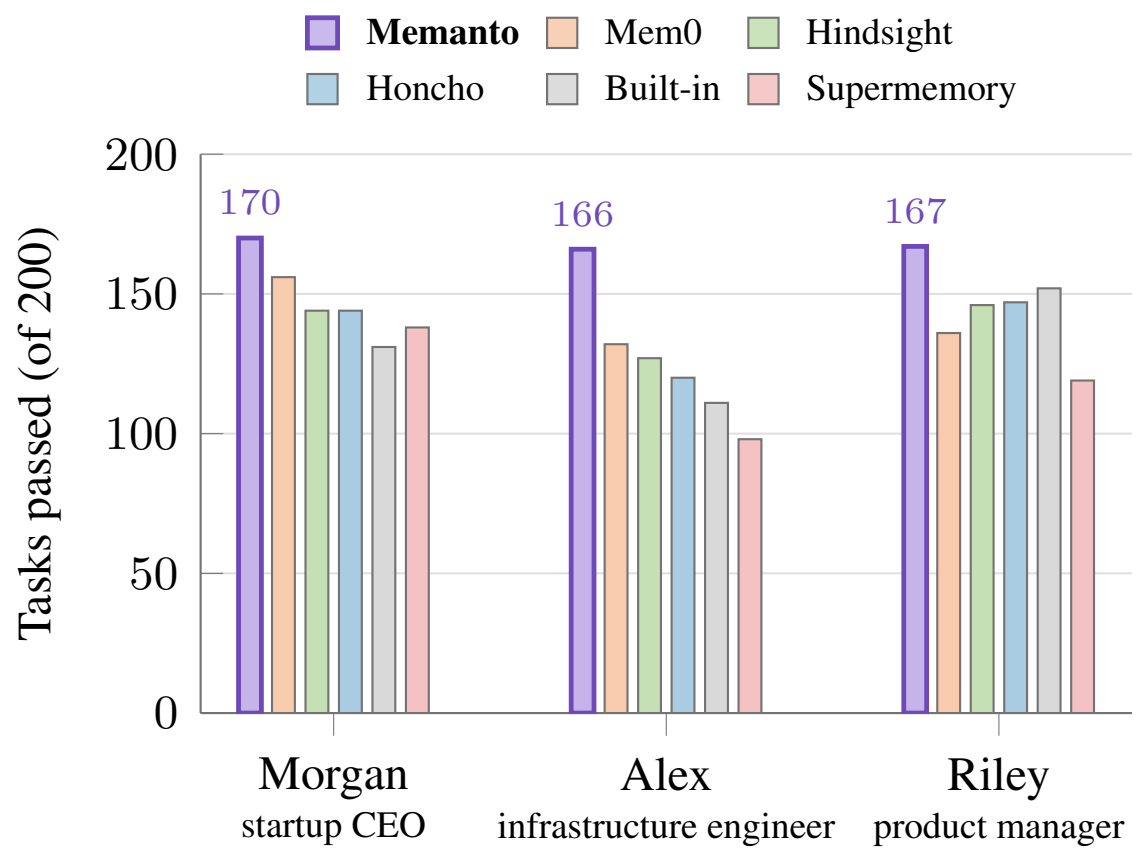

\paragraph{Personas.}
Figure~\ref{fig:persona} breaks the results down by persona, using the
per-persona scores in the official run records
\citep{dolphinbench2026results}. Memanto's run passes 170 of 200 tasks
for Morgan (85.0\%), 166 for Alex (83.0\%), and 167 for Riley (83.5\%):
four tasks separate its best and worst persona. The reported runs vary
far more, by 19 tasks (Hindsight) to 41 tasks (built-in memory). Alex,
the infrastructure engineer, is the hardest persona for every reported
run, which passes 98--132 of his tasks; Memanto's run passes 166, 34 more
than the best of them. On Morgan and Riley its margin over the best
reported run is 14 and 15 tasks. The agent model and settings differ
here as in Table~\ref{tab:results}.

\section{Discussion}
\label{sec:discussion}

\paragraph{Why retrieval may matter for action.}
One possible explanation for these results is how Memanto selects records
for the agent. Action tasks often depend on facts that change: a request
may depend on a preference stated in one interaction and revised several
sessions later. If retrieval returns only the earlier passage, the agent
can identify the right preference and still act on an outdated value. A
threshold that admits every memory not rated irrelevant, combined with a
generous limit, makes it more likely that both the original statement and
its revision reach the agent, which can then resolve the update before
acting. We have not isolated this effect; an ablation over the threshold
and the limit, and an analysis of the 97 failed tasks, would test it.

\paragraph{Where the cost and latency differences come from.}
In the reported Hermes runs, most of the agent cost was spent during
ingestion, when the agent processed the persona histories. The official
run records put agent inference for the 600 tests themselves at \$3.23
(Mem0) to \$8.05 (built-in memory), out of \$57.99--\$96.56 in agent
cost overall \citep{dolphinbench2026results}. In our run the agent also
processed every history message, yet its agent cost for ingestion and
tests together was \$2.17. Several factors contribute: the agent had a
single tool and a short prompt during ingestion; it used medium rather
than high reasoning effort and at most 6 rather than 60 iterations per
conversation; bounded retrieval (at most 50 memories) kept its context
small during tests; and GPT-6-Luna and GPT-5.6-Luna may differ in price
and speed. The same factors bear on latency. The iteration limit and
reasoning effort constrain the agent, so they are more likely to lower
accuracy than to raise it, yet the run completed more tasks than any
reported run. Whether the reported systems would keep their accuracy
under the same limits is untested; our data do not separate the effect
of memory from that of the model and settings, and a run of the
baselines with our settings, or of Memanto with theirs, would.

Memory-side cost depends less on the agent. Memanto's \$8.69 is the
Moorcheh Cloud usage credits the run consumed, against \$26.66--\$281.79
for the four external memory systems in the reported runs. A self-hosted
Moorcheh deployment would remove this usage charge, leaving only the cost
of running its infrastructure.

\section{Conclusion}

We evaluated Memanto on DolphinBench, where an agent must act on
information from earlier interactions without being told what to
retrieve. In the Hermes harness with GPT-6-Luna, Memanto completed 503 of
600 tasks (83.83\%) at a total cost of \$10.86 and a median task latency
of 15.32\,s, each better than every configuration reported by the
benchmark's authors. It did so with at most 6 agent iterations per
conversation, a tenth of those allowed in the reported runs. Because the
runs also differ in agent model, the comparison is between complete
systems; controlled runs with matched models and settings, ablations of
the retrieval settings, and repeated trials are the natural next steps. We release the scripts and results of this study at
\repourl\ so that others can repeat the run and carry out these
comparisons; at \$10.86 for a full run, doing so is inexpensive.
Benchmarks such as DolphinBench, which validate every
task against memory and require cost and latency alongside accuracy, make
such comparisons possible. They show that the choice of memory system
decides not only what an agent can recall, but whether it completes the
work, and at what cost.

\section*{Limitations}

\paragraph{Different agent model and settings.}
The reported baselines used GPT-5.6-Luna with high reasoning effort and
up to 60 iterations per conversation; our run used GPT-6-Luna with medium
reasoning effort. A different model may complete more
tasks, use fewer tokens, or respond faster regardless of the memory
system, and tighter limits lower cost and latency, so the gaps in
Table~\ref{tab:results} cannot be attributed to Memanto alone.

\paragraph{Cost accounting.}
Our agent cost is Hermes's own estimate from recorded API usage, with
retries not audited separately, and our memory cost is Moorcheh Cloud
usage credits. The reported costs follow the benchmark's own accounting,
so the two may differ in what they include.

\paragraph{Reported, not re-run, baselines.}
Baseline figures come from \citet{rathi2026dolphinbench}. Harness
versions, provider prices, and infrastructure may differ between their
runs and ours.

\section*{Ethical Considerations}

Several authors are affiliated with EdgeAI Innovations Inc., which
develops Memanto and the Moorcheh retrieval engine it uses.
To reduce the risk of bias, we use a public benchmark with fixed tasks,
the benchmark's own grading checks, prompts, and judge, and baseline
figures reported by the benchmark's authors rather than our own runs of
competing systems, and we state the differences in agent model and
settings throughout. So that our
results can be checked independently, we release the scripts and results
of this study in the accompanying repository.\footref{fn:repo}
DolphinBench personas are synthetic, and no personal data was used.

\bibliography{custom}

\end{document}